\documentclass[preprint]{aastex631}

\usepackage[utf8]{inputenc}
\usepackage{newunicodechar,graphicx}
\usepackage[normalem]{ulem}
\usepackage{hyperref}

\DeclareRobustCommand{\okina}{%
  \raisebox{\dimexpr\fontcharht\font`A-\height}{%
    \scalebox{0.8}{`}%
  }%
}
\newunicodechar{ʻ}{\okina}
\newcommand{\gaia}{\textit{Gaia} }
\newcommand{\gaias}{\textit{Gaia}}

\accepted{to \apjs}

\shorttitle{$H$ DIA light curves of MW Cepheids}
\shortauthors{Konchady et al.}

\graphicspath{{./}{figures/}}

\begin{document}

\title{$H$-band light curves of Milky Way Cepheids via Difference Imaging}

\correspondingauthor{Tarini Konchady}
\email{tkonchady@tamu.edu}

\author[0000-0003-0452-9182]{Tarini Konchady}
\affiliation{George P. and Cynthia W. Mitchell Institute for Fundamental Physics \& Astronomy,\\ Department of Physics \& Astronomy, Texas A\&M University, College Station, TX, USA}

\author[0000-0002-0582-1751]{Ryan J.~Oelkers}
\affiliation{George P. and Cynthia W. Mitchell Institute for Fundamental Physics \& Astronomy,\\ Department of Physics \& Astronomy, Texas A\&M University, College Station, TX, USA}
\affiliation{Vanderbilt University, Department of Physics \& Astronomy,  6301 Stevenson Center Ln., Nashville, TN 37235, USA}

\author[0000-0002-6230-0151]{David O.~Jones}
\affiliation{Department of Astronomy and Astrophysics, University of California, Santa Cruz, CA 95064, USA}
\affiliation{NASA Einstein Fellow}

\author[0000-0001-9420-6525]{Wenlong Yuan}
\affiliation{Department of Physics and Astronomy, Johns Hopkins University, Baltimore, MD 21218, USA}

\author[0000-0002-1775-4859]{Lucas M.~Macri}
\affiliation{George P. and Cynthia W. Mitchell Institute for Fundamental Physics \& Astronomy,\\ Department of Physics \& Astronomy, Texas A\&M University, College Station, TX, USA}

\author[0000-0001-8596-4746]{Erik R.~Peterson}
\affiliation{Department of Physics, Duke University, Durham, NC 27708, USA}

\author[0000-0002-6124-1196]{Adam G.~Riess}
\affiliation{Department of Physics and Astronomy, Johns Hopkins University, Baltimore, MD 21218, USA}
\affiliation{Space Telescope Science Institute, 3700 San Martin Drive, Baltimore, MD 21218, USA}

\begin{abstract}

We present $H$-band light curves of Milky Way Classical Cepheids observed as part of the DEHVILS survey with the Wide-Field Infrared Camera on the United Kingdom InfraRed Telescope. Due to the crowded nature of these fields caused by defocusing the Camera, we performed difference-imaging photometry by modifying a pipeline originally developed to analyze images from the Transiting Exoplanet Survey Satellite. We achieved a photometric precision in line with expectations from photon statistics, reaching 0.01~mag for $8 \lesssim H \lesssim 11$~mag. We used the resulting Cepheid light curves to derive corrections to ``mean light'' for random-phase {\it Hubble Space Telescope} observations in {\it F160W}. We find good agreement with previous phase corrections based on $VI$ light curves from the literature, with a mean difference of $-1\pm 6$ millimag.

\end{abstract}

\section{Introduction} \label{sec:intro}

For over a hundred years, classical Cepheid variables (hereafter Cepheids) have been known to follow Period--Luminosity relations  \citep[PLRs;][]{leavitt1908,leavitt1912}. This property has made Cepheids crucial for establishing the Extragalactic Distance Scale and consequently measuring the Hubble constant (H$_0$). Near-infrared (NIR) observations of Cepheids have proven especially useful for distance measurements as the intrinsic width of the PLR is only $\sim 0.08$ mag at those wavelengths \citep{persson2004,macri2015,riess2019} in addition to being less sensitive to dust and metallicity.

Interestingly, local measurements of H$_0$ based on Cepheids and type Ia supernovae \citep[SNe\,Ia;][]{riess2019,riess2021} differ from expectations based on the Lambda Cold Dark Matter ($\Lambda$CDM) cosmological model, anchored by observations of the Cosmic Microwave Background \citep{planck} and Baryon Acoustic Oscillations \citep{bao}, at the $\sim4-6\sigma$ level. The tantalizing hint of ``New Physics" that could be the cause of  this discrepancy requires identifying and addressing any systematic uncertainties in the respective measurements.

One source of systematic uncertainty in local measurements of H$_0$ comes from comparing Milky Way (MW) Cepheids to their extragalactic counterparts. One route to addressing this uncertainty lies in the \gaia mission \citep{gaia2016a}, which is providing high-precision parallaxes for billions of objects, including MW Cepheids. Presently, the intermediate \gaia data products require parallax offsets that depend on magnitude and ecliptic latitude \citep{lindegren2018,lindegren2021a,lindegren2021b}. 

\textit{Hubble Space Telescope} ({\it HST}) observations of MW Cepheids can play a critical role in local H$_0$ measurements, as they can be simultaneously used to determine the \gaia parallax offset for Cepheids and reduce the zeropoint errors that arise while comparing MW Cepheids to extragalactic variables. Such observations were obtained for {29/40 targets} during {\it HST} Cycle 27  ({prop \#15879, PI Riess}). {The targets were all known MW Cepheids with photometric parallaxes $\pi_{\rm phot}\!>\!0.8$~mas (to maximize \\sensitivity to the \gaia offset), $V\!>\!6$~mag (to avoid saturation by \gaias) and $A_H\!<\!0.6$~mag (to minimize the impact of reddening). Since the {\it HST} observations were obtained} at random phases due to the ``snapshot'' nature of {the} program{, g}round-based optical or NIR light curves must be used to correct the single-epoch {\it HST} magnitudes to ``mean light,'' or the mean {\it HST} magnitude of a given Cepheid. {\citet{riess2021} presented results based on all observations obtained for this program through the end of 2020 (25/29 targets), relying on {\it VI} data for phase corrections. They obtained H$_0=73.2\pm1.3$~km/s/Mpc, a 1.8\% measurement that exceeds the Planck CMB+$\Lambda$CDM expectation by $4.2\sigma$.}

{As part of this {\it HST} program, we aimed to obtain $H$-band ground-based light curves for as many of these Cepheids as possible to check the consistency of phase corrections obtained from optical and NIR data. We used} observations taken for the Dark Energy, H$_0$, and peculiar Velocities using Infrared Light from Supernovae (DEHVILS) survey using the Wide-Field Infrared Camera (WFCAM) on the United Kingdom InfraRed Telescope (UKIRT). To avoid saturation from the target Cepheids, WFCAM is defocused prior to the Cepheid observations. Consequently, the resulting images show significant crowding and blending. This renders typical aperture or point-spread function (PSF) photometry techniques less effective, which is why we turned to difference imaging analysis (DIA).

The basis of DIA is determining the flux difference of a source between a reference and a science image. The reference image typically has the highest signal-to-noise ratio (and in some cases the best resolution) of all available images and is degraded to match the conditions of a given science image before it is subtracted from the science image. Objects with constant flux levels will be subtracted into background noise while variable objects will leave behind some residual flux \citep{alard1998}. Aperture or PSF photometry can then be performed on the differenced images to extract the light curves of variable objects. {The DIA implementation used in this work is a slightly modified version of the method presented in \citet{oelkers2018} and \citet{oelkers2019}, which was developed to extract light curves from images taken by the Transiting Exoplanet Survey Satellite (\textit{TESS}).}

{The rest of this paper is organized as follows:} \S\ref{sec:obs} describes the DEHVILS survey along with the $H$-band Cepheid observations and image preprocessing, \S\ref{sec:dia} describes the DIA procedure, \S\ref{sec:light_curves} presents the Cepheid light curves and compares the derived phase corrections to {\it HST} observations to similar corrections based on $VI$ light curves, and \S\ref{sec:summary} provides a summary of this work.

\section{Observations and Image Preprocessing} \label{sec:obs}

\subsection{The DEHVILS Survey} \label{sec:dehvils}

The DEHVILS survey started in northern Spring 2020 with the primary goal of using UKIRT to build a NIR sample of SNe\,Ia.  The survey aims to measure the local growth of structure parameters and provide an ``anchor" sample for next-generation high-redshift samples such as those from the Vera C.~Rubin Observatory and the {\it Nancy Grace Roman Space Telescope}.  DEHVILS has observed over 100 SNe\,Ia in $YJH$ in its first year of operations and, with collaborators at the University of Hawai\okina i, over 300 SNe\,Ia in $J$.

\begin{figure}[t]
\plotone{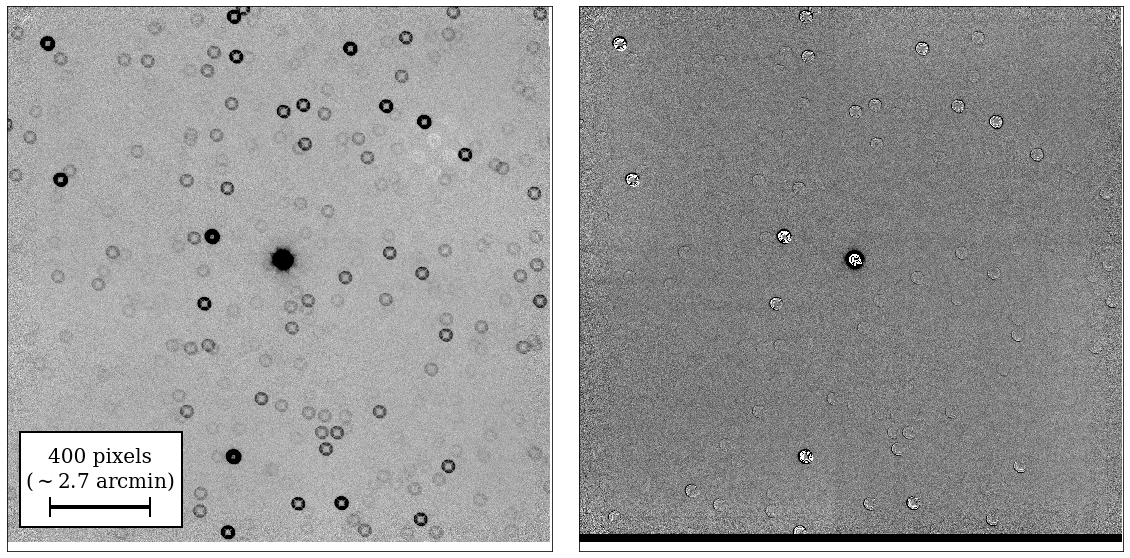}
\caption{A representative science image (left) of the RX Cam field, with the Cepheid near the center. The differenced image (right) is the result of subtracting the convolved reference image from the science image. The color scale is inverted for clarity. North is up and east is to the left.}
\label{fig:wfcam_rxcam_field}
\end{figure}

UKIRT's WFCAM has already observed $\sim$17,900 square degrees of sky in $J$ as part of the UKIRT Hemisphere Survey \citep{Dye18} and $\sim$6,200 square degrees in $zYJHK$ as part of the UKIRT Infra-red Deep Sky Survey \citep[UKIDSS;][]{Lawrence07}.  Thanks to the large-area footprint and WFCAM's $\sim$1-degree field of view, the photometry can be calibrated relative to the 2MASS Point Source Catalog \citep[PSC,][]{2MASS} at the $\sim 1\%$~level \citep{Hodgkin09}.  DEHVILS also uses observations of {\tt CALSPEC} standard stars \citep{Bohlin96} with in-focus and defocused observations to measure the calibration and linearity of the UKIRT system.

\subsection{Observations \& Image Preprocessing}

{DEHVILS targeted} 12 MW Cepheids {from the aforementioned {\it HST} program that were observable from Hawai\okina i.} Though these Cepheids will normally saturate at the minimum UKIRT exposure {times}, by defocusing the telescope we can avoid the non-linear regime. {We present the analysis for seven of these variables} whose observations have been completed. Each target was observed for 11--20 epochs spread over 2--3 months between May and October 2020.

{Table \ref{table:mw_cep_table} presents the mean $H$ magnitudes and periods of these objects. The magnitudes and associated uncertainties are from the 2MASS Point Source Catalog (PSC; \citealt{2MASS}) while the periods are from \citet{riess2021}. The Cepheid periods were derived while applying the phase correction procedure that is described briefly in Section \ref{subsec:phase} and in more detail in the appendix of \citet{riess2018b}. The procedure relies on multiband literature photometry, whose sources are shown in Tables 3 and 4 of \citet{riess2021}. The long baseline of the photometry used ($\sim$20--25 years) yields negligible uncertainties in the derived periods.}

WFCAM is a wide-field infrared camera consisting of four detectors (arranged in a $2\times 2$ array), each with a field of view of 0.21 square degrees and a plate scale of $0.4\arcsec$ per pixel. The detectors cover $13.65\arcmin$ on a side and are spaced $12.65\arcmin$ apart. Available filters include {\it zYJHK} \citep{wfcam}.

In a given exposure, the relevant Cepheid was observed by one of the four WFCAM detectors. We obtained two images per epoch for a given Cepheid, rotated 90 degrees from each other. The left panel of Figure~\ref{fig:wfcam_rxcam_field} shows a typical image of one of our targets. 

Images of a given Cepheid were aligned by updating their WCS information prior to running the DIA pipeline. The first image from the first epoch of a given object was adopted as the reference WCS. At least twelve bright, isolated stars were identified in all images of a given field and their $(x,y)$ positions were used to derive geometric transformations using IRAF\footnote{Image Reduction and Analysis Facility \citep{Tody1986}.}, with iterative rejection of outliers. 

\begin{deluxetable}{llrrr}[t]
\tablecaption{Milky Way Cepheids observed by the DEHVILS survey}
\label{table:mw_cep_table}
\tablehead{\multicolumn{1}{l}{Name} & \colhead{RA}  & \colhead{Dec} & \colhead {$H$}  & \colhead{log $P$}\\
 \colhead{} & \multicolumn{2}{c}{(J2000)} & \colhead{[mag]}& \colhead{[day]}}
\startdata
RX Cam    & 04:04:58.5  & +58:39:35.2   & 4.864 $\pm$ 0.021 & 0.898 \\
RV Sco    & 16:58:19.7  & $-$33:36:32.8   & 4.817 $\pm$ 0.075 & 0.783 \\
BF Oph    & 17:06:05.5  & $-$26 34 50.0   & 5.282 $\pm$ 0.043 & 0.609 \\
AP Sgr    & 18:13:02.5  & $-$23:07:02.2   & 4.980 $\pm$ 0.039 & 0.704 \\
SS Sct    & 18:43:43.5  & $-$07:43:52.0   & 5.910 $\pm$ 0.031 & 0.565 \\
TX Cyg    & 21:00:06.4  & +42:35:51.2   & 4.844 $\pm$ 0.025 & 1.168 \\
V0386 Cyg\ \ \ \ \  & 21:14:40.4  & +41:42:58.8   & 5.700 $\pm$ 0.009 & 0.721 
\enddata
\tablecomments{$H$ magnitudes and uncertainties from the 2MASS PSC \citep{2MASS}; periods from \citet{riess2021}.}
\end{deluxetable}

\section{Difference Imaging Procedure} \label{sec:dia}

\subsection{DIA Pipeline for \textit{TESS} Full-Frame Images}
This work used a modified version of the difference imaging pipeline from \citet[OS-DIA]{oelkers2015, oelkers2018, oelkers2019} to measure the photometry of each Cepheid. The OS-DIA pipeline was originally designed to measure stellar photometry from defocused images generated by the Chinese Small Telescope Array (CSTAR), and was adapted to extract light curves from \textit{TESS} full-frame images (FFIs) \citep{oelkers2015, oelkers2018}. The pipeline reduced more than $10^6$ images from CSTAR, and has generated more than 100 million light curves from \textit{TESS} FFIs with a precision that has met the expectation of initial prediction models (60 ppm hr$^{-0.5}$; \citealt{ricker2014}, \citealt{sullivan2015}; \citealt{oelkers2015})

\begin{figure}[t]
\begin{center}
\includegraphics[width=0.8\textwidth]{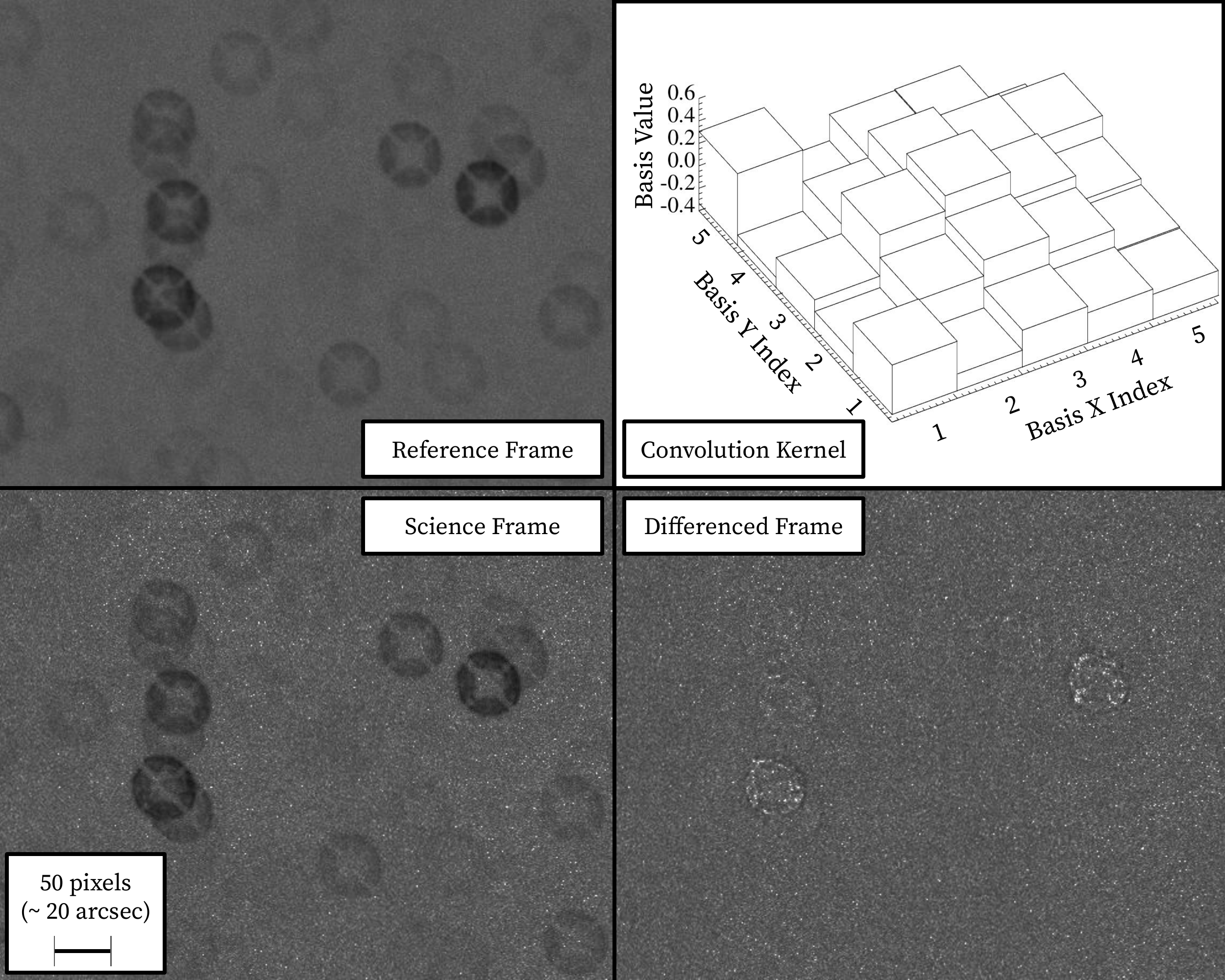}
\end{center}
\caption{Top-left: a $500\times400$ pixel cutout of the RX~Cam reference frame. Top-right: a 3-D visualization of the $5\times 5$ $\delta$-function kernel used to convolve the reference frame (top-left) to match a typical science frame (bottom-left). Bottom-left: cutout of the same area for a typical science frame in the RX~Cam image sequence. Bottom-right: residuals after subtraction. All images displayed using the same logarithmic stretch; colors have been inverted for clarity.}
\label{fig:kernel}
\end{figure}

The OS-DIA pipeline uses a Dirac $\delta$-function kernel to transform reference images and account for ``non-Gaussian, arbitrarily-shaped PSFs", such as those seen in the defocused WFCAM images of this work. This kernel type provides more flexibility when characterizing non-Gaussian PSFs because each individual kernel basis is independently solved for, which results in a kernel-map that is not required to be Gaussian in shape. Light curves of all objects are extracted from the differenced images via aperture photometry and detrended using the light curves of sources with low dispersion that have similar magnitudes and are nearby to the variable objects on the detector. We employed the OS-DIA pipeline on the WFCAM images with a spatially constant $5\times5$ pixel kernel after our initial testing showed first- and second-order spatially varying kernels provided little improvement in photometric precision but significantly increased the runtime of the pipeline. {Figure~\ref{fig:kernel} shows a typical kernel for one of our images and its efficacy at convolving the reference image to match the PSF of the image to be differenced.}

\subsection{Pipeline Modifications for WFCAM Images}

We made a number of modifications to the OS-DIA pipeline to improve its performance on our defocused images. First, we used the coordinates of the center of the reference image to query the \textit{TESS} Input Catalog \citep[TIC]{stassun2018} to get a list of sources that appeared in the image. However, since the WFCAM images were defocused, there was a consistent offset between the TIC coordinates and the WFCAM initial WCS. We removed this offset by visually identifying the stellar centroids {that would capture the stellar flux completely within our photometry apertures, which varied between 26-41 pixels depending on the defocused nature of the Cepheid.}

We median-combined all the images of a given Cepheid to generate the reference frame used for subtraction. This differs from the procedure in \citet{oelkers2018,oelkers2019} which only used the first image in the series as the reference image.

We modified the selection procedure to identify stars which could be used to solve for the reference kernel. We only selected stars which had pixel positions farther than at least 100 pixels from the edge, and had photometric uncertainties less than 0.05~mag after an initial execution of aperture photometry on the reference frame. Additionally, we purposefully excluded the Cepheid from the list of stars that could be used for the kernel generation since its variability would likely degrade the kernel solution.

Finally, we modified the original OS-DIA light curve detrending procedure applied to the Cepheids. Normally, this pipeline uses a median-combined subset of 100 stars of similar magnitude to the target star which decreases the photometric dispersion when combined and subtracted from the target light curve. This traditional approach was used to evaluate our photometric precision as discussed in \S\ref{sec:photprec}. However, we were unable to use this method for the Cepheids as there are few (if any) stars with similar magnitudes in each frame. Instead, we selected all stars within 250 pixels of the Cepheid as ``trend" stars. Next, we subtracted the reference frame magnitude of each trend star from its full light curve, and median-combined the trend light curves with a $2\sigma$ clipping to create a reference trend. This reference trend was then subtracted from the light curve of the corresponding Cepheid.

\subsection{Evaluating the Photometric Precision of Output Light Curves \label{sec:photprec}}

\begin{figure}[t]
\begin{center}
\includegraphics[height=5in]{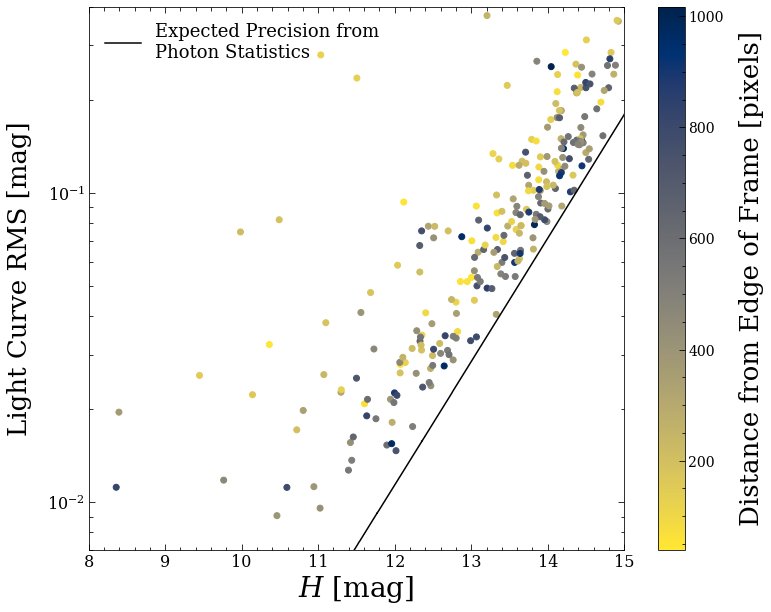}
\end{center}
\caption{Photometric precision of light curves obtained via difference imaging (colored points) and the expected precision from photon statistics (solid line) for a representative field (RX Cam). The color of each point indicates distance from the edge of the frame, showing greater precision in the central area.}
\label{fig:noise_floor}
\end{figure}

We characterized the photometric quality of the differenced light curves for stars other than the Cepheids as follows. We first subtracted the mean magnitude of each star in every field from the corresponding detrended light curve and then computed the median absolute value of the resulting offsets, performing iterative $5\sigma$ clipping to exclude outliers. The results for one representative field are shown in Figure \ref{fig:noise_floor}. We achieved a photometric precision limit of $\sim 0.01$~mag for bright ($8 \lesssim H \lesssim 11$~mag) stars. We investigated whether the achieved photometric precision was in line with expectations by performing aperture photometry on the raw images of the RX Cam field using the corresponding input star list to the pipeline. We determined the $S/N$ of each object taking into account contributions from photon statistics, sky background, and readout noise. As shown by the solid black line in Figure \ref{fig:noise_floor} the photometric precision expected from $S/N$ considerations closely follows the noise floor. The objects with excess r.m.s. ($0.07-0.2$~mag for $11.5 < H < 13.3$) are either uncharacterized variables or located near the edges of the reference image, where the quality of the image subtraction and subsequent photometry procedures are less reliable.

\clearpage

\begin{deluxetable}{lcc}[b]
\tablecaption{Parameters from Template Fitting of Cepheid Light Curves}
\label{table:lc_fit_params}
\tablehead{\colhead{Name}\hspace{0.75in} & \colhead{\hspace{0.75in}Amplitude}\hspace{0.75in} & \colhead{Phase offset}\hspace{1in}}
\startdata
AP Sgr    & 0.238 $\pm$ 0.013 & 0.032 $\pm$ 0.008 \\
BF Oph    & 0.237 $\pm$ 0.017 & 0.084 $\pm$ 0.009 \\
RV Sco    & 0.226 $\pm$ 0.015 & 0.041 $\pm$ 0.008 \\
RX Cam    & 0.178 $\pm$ 0.022 & 0.049 $\pm$ 0.014 \\
SS Sct    & 0.192 $\pm$ 0.005 & 0.030 $\pm$ 0.005 \\
TX Cyg    & 0.285 $\pm$ 0.020 & -0.007 $\pm$ 0.011 \\
V0386 Cyg & 0.259 $\pm$ 0.010 & 0.042 $\pm$ 0.004
\enddata
\end{deluxetable}
\clearpage

In the case of the Cepheids, we phased their light curves adopting the periods listed in Table~\ref{table:mw_cep_table} and fit them using templates from \citet{Inno2015}{\footnote{The template-fitting code is available at \href{https://github.com/wenlong2/Fit2Inno2015}{github.com/wenlong2/Fit2Inno2015}}, which is linear in amplitude and mean magnitude, and nonlinear in initial phase offset. We adopted a strategy of first searching for the initial phase offset that achieved a global least-squares minimum, simultaneously solving for amplitude and mean magnitude for each trial value of the initial phase. Then, the initial phase offset and parameter uncertainties were fine-tuned using the Gauss-Newton algorithm. This strategy ensures both accuracy and speed. The best-fit model amplitudes and initial phase offsets, along with their uncertainties, are listed in Table \ref{table:lc_fit_params}.}

We used the residuals {from the light curve fitting} to estimate a global statistical uncertainty of 0.027~mag for the Cepheid photometry. This larger value relative to the brightest non-Cepheids in the frame likely arises from our limited ability to detrend the former light curves. Overall, the phase correction uncertainties are dominated by the light curve modeling errors, and thus were estimated by the scatter of the light curve fitting residuals.

\section{Cepheid light curves and Phase Corrections} \label{sec:light_curves}

\begin{figure}[t]
\begin{center}
\includegraphics[height=7.25in]{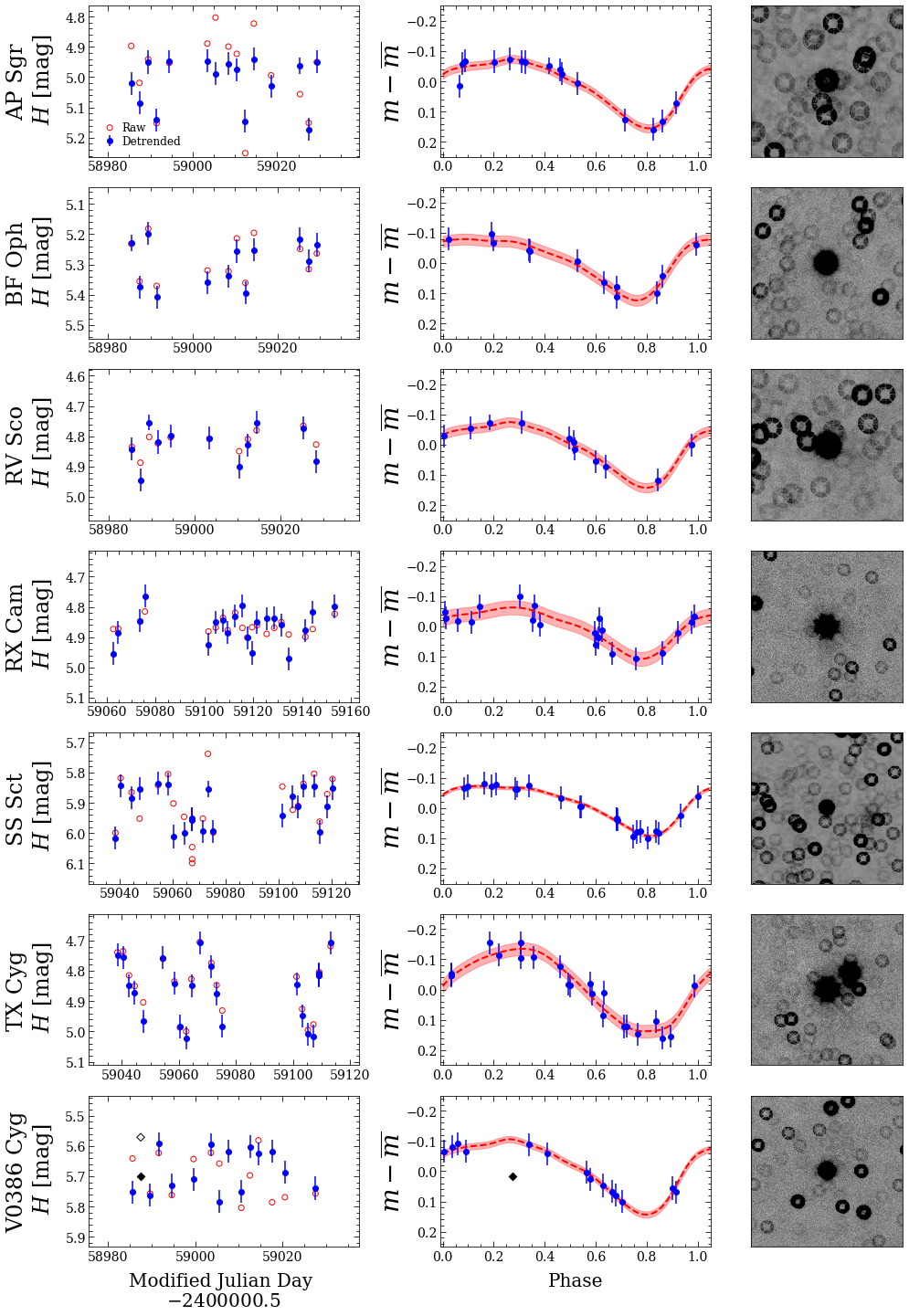}
\end{center}
\caption{Raw (open red {circles}) and detrended (left) and phased (center) light curves (filled blue {circles}) and models (dashed red lines), plus reference images (right, $4\arcmin$ on a side, inverted for clarity) of our target Cepheids. The mean magnitudes used in the left column were taken from 2MASS PSC. Phased light curves are plotted relative to the mean magnitude of each light curve{, and the models are shown with $1\sigma$ intervals (shaded red regions)}. Observations taken within 2.4~h were averaged into a single epoch. Periods were taken from \citet{riess2021}. The black {diamonds} in the V0386 Cyg light curves correspond to an epoch that was excluded due to a significant difference in defocused PSF size compared to all other images.}
\label{fig:cep_lcs}
\end{figure}

Table~\ref{table:phot} presents our fully-calibrated photometric measurements; observations taken within 2.4 hours were averaged into a single epoch. Figure~\ref{fig:cep_lcs} shows the raw, detrended, and phased Cepheid light curves and also includes ``postage stamps'' of $4\arcmin$ around each variable.

\subsection{Comparison of Derived Cepheid Amplitudes with Previous Studies}

\begin{figure}[t]
\begin{center}
\includegraphics[width=0.9\textwidth]{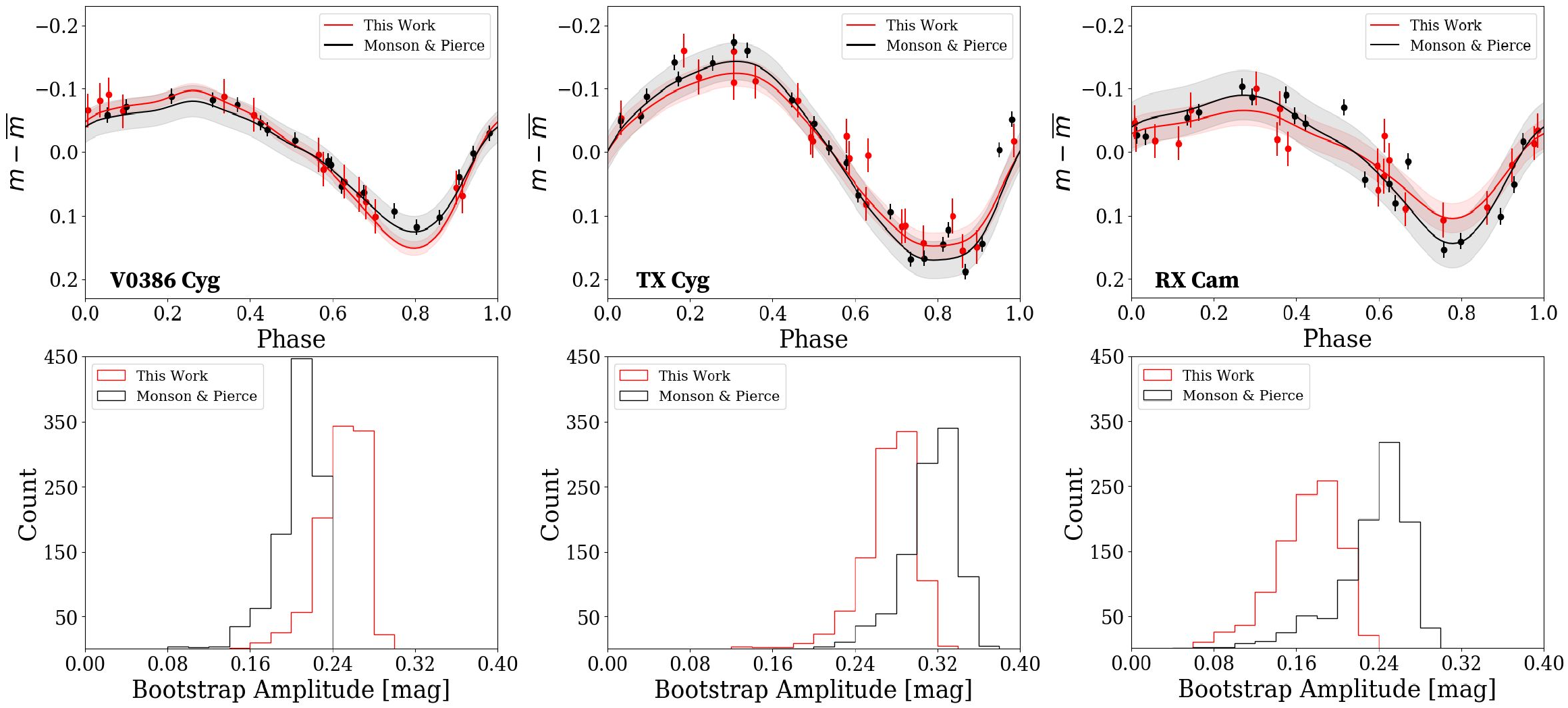}
\end{center}
\caption{Results of the bootstrap procedure to compare Cepheid amplitudes derived from our work (red) and from \citet[][black]{monson2011}. Top: Binned and phase-folded light curves; solid lines show the best-fit templates. {The red and black shaded regions represent the $1\sigma$ intervals}. Bottom: Results of 1000 bootstrap samplings (with replacement) to estimate the uncertainty on the best-fit amplitudes. We find agreement between the two data sets at the $1.1-1.4\sigma$ level. }
\label{fig:bootstraps}
\end{figure}

We compared three of our Cepheid light curves (V0386 Cyg, TX Cyg, and RX Cam) with those obtained by \citet{monson2011} to provide context into our template fitting and data reduction. We executed a bootstrap simulation sampling from both sets of light curves independently (with replacement) 1000 times. We scaled the amplitude of the Cepheid template (described in \S\ref{subsec:phase}) during each bootstrap simulation and selected the amplitude which minimized the least-squares residuals of the fit.  The results are presented in Figure~\ref{fig:bootstraps}.

{We found the amplitudes for V0386 Cyg to be 0.25~mag in this work and 0.21~mag from \citet{monson2011}, which are consistent within $1.0\sigma$ using the photometric uncertainties of the light curves and $1.3\sigma$ using the standard deviation of the bootstrap simulations. We found the amplitudes for TX Cyg to be 0.27~mag in this work and 0.31~mag from \citet{monson2011}, which are consistent within $1.0\sigma$ using the photometric uncertainties of the light curves and $1.0\sigma$ using the standard deviation of the bootstrap simulations. Lastly, we found the amplitudes for RX Cam to be 0.17~mag in this work and 0.23~mag from \citet{monson2011}, which are consistent within $1.5\sigma$ using the photometric uncertainties of the light curves and $1.4\sigma$ using the standard deviation of the bootstrap simulations.} We interpret these results as being statistically consistent. 

\clearpage

\begin{deluxetable}{lrrrr}[t]
\tablecaption{Cepheid Photometry}
\label{table:phot}
\tablehead{\multicolumn{1}{l}{Name\hspace{1in}} & \multicolumn{1}{c}{\hspace{1.25in}MJD$^a$\hspace{1.25in}} & \multicolumn{1}{c}{\hspace{1.25in}Phase$^b$\hspace{1.25in}} & \multicolumn{1}{c}{\hspace{1.25in}$H$ [mag]$^c$\hspace{1.25in}}}
\startdata
AP Sgr & 8985.4918 & 0.529 &  5.024\\
BF Oph & 8985.4344 & 0.197 &  5.228\\
RV Sco & 8985.4264 & 0.516 &  4.858\\
RX Cam & 9062.5855 & 0.665 &  4.941\\
SS Sct & 9038.4188 & 0.805 &  6.023\\
TX Cyg & 9038.4533 & 0.221 &  4.695\\
V0386 Cyg & 8985.6038 & 0.915 &  5.761
\enddata
\tablecomments{(a) JD$-2450000.5$. (b) based on the periods listed in Table~\ref{table:mw_cep_table} and the phase offsets listed in Table~\ref{table:lc_fit_params}; the overall systematic uncertainty in this parameter for a given Cepheid is provided in the latter table. (c) DIA magnitude + mean 2MASS magnitude from Table~\ref{table:mw_cep_table}; a statistical uncertainty of 0.027~mag applies to all lines (see \S\ref{sec:photprec}). Only a few lines are shown here for guidance; the full version is available as a machine-readable file.}
\end{deluxetable}

\subsection{Comparison of Phase Corrections Based on \textit{VI}- and \textit{H}-band light curves \label{subsec:phase}}

We obtained corrections to ``mean light'' for the random-phase {\it HST} Wide Field Camera 3 (WFC3) $F160W$ observations of these Cepheids reported in \citet{riess2021}. One set of corrections was based on $V$- and $I$-band light curves from the literature (see Tables 3 and 4 in \citealt{riess2021}) while the other was based on our $H$-band light curves. The procedure to obtain the phase corrections is described in detail in the appendix of \citet{riess2018a}, but we briefly summarize the procedure below. 

First, any available observations in the $VIJH$ bands that contain epochs close to the {\it HST} observations of a given Cepheid are assembled. These bands are used because they are similar in central wavelength and bandpass to the {\it HST} WFC3 bands $F555W$, $F814W$ and $F160W$, and thus can be easily transformed. The assembled observations are then combined into a single dataset, which is fit with a Fourier series to obtain a model of Cepheid variability. \citet{riess2018a} consider two models of variability: one where the period is kept constant and another where the period is allowed to vary along with the other model parameters. For the set of phase corrections presented here, the constant period model was used.

The variability model is then used to convert observation times to phase, and a cubic spline (or a Cepheid template if the number of observations is limited) is used to interpolate the light curves in a single band and determine magnitude at the observed phase, $m_\phi$. $m_\phi$ is used to define a phase-correction curve $C_\phi = \overline{m} - m_\phi$. The $H$-band phase was allowed to vary freely and was not shifted relative to the $V$-band phase. Photometric transformations from \citet{riess2016} are then used to convert the phase corrections from the ground-based to the {\it HST} photometric system. 

\begin{figure}[b]
\begin{center}
\includegraphics[height=4in]{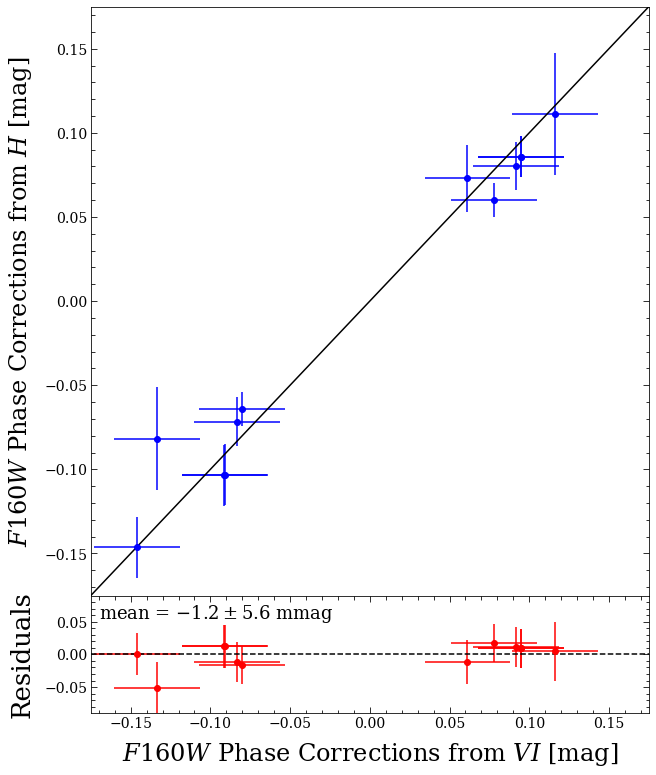}
\end{center}
\caption{Comparison of corrections to ``mean light'' for the random-phase {\it HST F160W} observations of the seven MW Cepheids discussed in this work, derived from ground-based $VI$ and $H$ light curves, respectively.}
\label{fig:phase_cors}
\end{figure}

Figure~\ref{fig:phase_cors} compares the two sets of phase corrections. We find good agreement with the phase corrections based on $VI$ light curves, with a small mean difference of $-1\pm 6$ millimag.

\clearpage

\clearpage

\clearpage

\section{Summary} \label{sec:summary}

We presented $H$-band light curves of seven MW Cepheids observed as part of the DEHVILS survey. We extracted the light curves using a modified difference imaging pipeline that has been recently adapted to \textit{TESS} FFIs and modified to account for the defocused observing mode. We find our adapted pipeline has achieved a photometric precision limit of $\sim 0.01$~mag. We used the resulting light curves to determine phase corrections for {\it HST} $F160W$ observations of these Cepheids and correct the measurements to ``mean light.'' We compared the $H$-band phase corrections to those obtained using $VI$-band light curves from the literature and found good agreement, with a mean difference of $-1\pm 6$ millimag.

\begin{acknowledgments}

We thank Watson Varicatt for providing the raw WFCAM images used for noise estimation. The original \textit{TESS} DIA pipeline was built on work from \citet{alard2000}, \citet{alard1998}, \citet{miller2008}, and \citet{oelkers2015}. UKIRT is owned by the University of Hawai\okina i (UH) and operated by the UH Institute for Astronomy. When (some of) the data reported here were obtained, the operations were enabled through the cooperation of the East Asian Observatory. This research makes use of data products from the Two Micron All Sky Survey, which is a joint project of the University of Massachusetts and the Infrared Processing and Analysis Center/California Institute of Technology, funded by the National Aeronautics and Space Administration and the National Science Foundation. Support for this work was provided by NASA through the NASA Hubble Fellowship grant HF2-51462.001 awarded by the Space Telescope Science Institute. This research is also based on observations made with the NASA/ESA {\it Hubble Space Telescope} obtained from the Space Telescope Science Institute, which is operated by the Association of Universities for Research in Astronomy, Inc., under NASA contract NAS 5–26555. This research has made use of NASA’s Astrophysics Data System Bibliographic Services, and the SIMBAD database, operated at CDS, Strasbourg, France.
\end{acknowledgments}

\facilities{UKIRT, {\it HST}}

\software{
    Astropy \citep{astropy1,astropy2}, 
    Project Jupyter \citep{jupyter}, 
    Matplotlib \citep{matplotlib}, 
    Numpy \citep{numpy}, 
    Pandas \citep{pandas}, 
    Photutils \citep{photutils}, 
    IRAF \citep{Tody1986},
    SAOImage DS9 \citep{saods9}.
          }

\clearpage
\bibliography{mwceph}{}
\bibliographystyle{aasjournal}
\end{document}